\documentclass[twocolumn,showpacs,preprintnumbers,prC]{revtex4}
\usepackage{graphicx}
\def\bea {\begin{eqnarray}}
\def\eea {\end{eqnarray}}
\def\ra {\rightarrow}
\def\be {\begin{equation}}
\def\ee {\end{equation}}
\def\nn {\nonumber}
\begin{document}
\title{\bf Momentum dependence of drag coefficients and heavy flavour suppression
in quark gluon plasma}
\author{Surasree Mazumder\footnote{surasree@vecc.gov.in}, Trambak Bhattacharyya, Jan-e Alam  and 
Santosh K Das}
\medskip
\affiliation{Theory Division, 
Variable Energy Cyclotron Centre, 1/AF, Bidhan Nagar ,
Kolkata - 700064}
\date{\today}
\begin{abstract}
The momentum dependence of the drag coefficient of heavy quarks
propagating through  quark gluon plasma (QGP) has been evaluated.
The  results have been used to estimate the nuclear suppression 
factor of charm and bottom quarks in QGP.  We observe that the 
momentum dependence of the transport coefficients plays crucial 
role in the suppression of the heavy quarks and consequently 
in discerning the properties of QGP using heavy flavours as 
a probe.  We show that the large suppression of the 
heavy quarks observed at RHIC and LHC  is predominantly due to the 
radiative losses. The suppression of  $D^0$  in Pb+Pb collisions 
at LHC energy - recently measured by the ALICE collaboration has also
been studied.
\end{abstract}

\maketitle

\section{\bf INTRODUCTION}
Simulations of QCD equation of state (EoS) on lattice 
show that at very high temperatures and/or densities
the nuclear matter undergoes
a phase transition to a new state of matter called Quark-Gluon Plasma(QGP). 
It is expected that QGP can be produced experimentally  by colliding
two nuclei at ultra-relativistic energies. Relativistic Heavy Ion 
Collider(RHIC) at BNL and Large Hadron Collider(LHC) at
CERN are two such experimental facilities.
The depletion of  hadrons with high transverse momentum ($p_T$)
produced in Nucleus + Nucleus collisions with respect to those
produced in proton + proton (pp) collisions has been considered
as a signature of QGP formation.
The two main processes which cause the depletion are
(i) the elastic collisions
and (ii) the radiative loss or the inelastic collisions 
of the  high energy partons with the quarks,
anti-quarks and gluons in the thermal bath.
The importance of elastic energy loss in QGP diagnosis was
pointed out first by Bjorken~\cite{bjorken}. 
Later the calculations of elastic loss  had been 
performed with improved 
techniques~\cite{TG,peshier}
and its importance was highlighted subsequently~\cite{akdm,MT}.

The abundance of charm and bottom quarks in the partonic plasma, 
in the expected range of temperature to be attained in the experiments, 
is small. Consequently, the bulk properties of the plasma is not decided 
by them and hence heavy quarks may act as an efficient probe for the 
diagnosis of QGP. 
The collisional energy loss of heavy quarks~\cite{braaten}
has gained importance recently in view of the measured
nuclear suppression in the $p_T$ spectra of 
non-photonic single electrons~\cite{stare,phenixe}. 
In the present work we focus on the energy loss of 
heavy quarks in QGP in deducing the properties of the medium. 

Several ingredients like inclusions of non-perturbative
contributions from the quasi-hadronic bound state~\cite{hvh},
3-body scattering effects~\cite{ko},
the dissociation of heavy mesons due to its 
interaction with the partons in the
thermal medium~\cite{adil}  and employment of running coupling
constants and realistic Debye mass~\cite{gossiaux}
have been proposed to improve the description of the  experimental data.
For mass  dependence of energy loss due to radiative processes 
Dokshitzer and Kharzeev~\cite{DK} argue that
radiative energy loss of heavy quarks
will be suppressed  compared to that of light quarks due to 
dead cone effects~\cite{rkellis}.  
However, Aurenche and Zakharov claim that the radiative process has an 
anomalous mass dependence~\cite{zakharov} due to the finite size of the QGP 
which leads to  small difference in energy loss between
a heavy and a light quarks.  
Although the authors of~\cite{roy} concluded that 
the suppression of radiative loss for heavy quarks is 
due to dead cone effects but it will be fair to state
that the issue is not settled yet.
 
The other mechanism that can affect the radiative loss is the 
LPM effect~\cite{wg} which 
depends on the relative magnitude of two time
scales of the system~\cite{klein}: the formation time  
($\tau_F$) and the mean scattering time scale ($\tau_{sc}$) 
of the emitted gluons.
If $\tau_F\,>\,\tau_{sc}$ then LPM suppression
will be effective.  The LPM effect is built-in in the expression for
radiative energy loss of heavy quarks derived in~\cite{asw,wang,adsw,dg}.
In the present work  
the LPM and the dead cone effects are explicitly
taken into account for heavy quark dissipation.

We assume here that  the light quarks and gluons thermalize 
before heavy quarks.  The charm and bottom quarks
execute Brownian motion~\cite{moore,japrl,sc,svetitsky,rapp,turbide,
bjoraker,npa1997,rma,dasetal,alberico} 
in the heat bath of QGP.
Therefore, the interaction of the heavy quarks with 
QGP may be treated as the 
interactions between equilibrium and non-equilibrium degrees
of freedom. 
The Fokker-Planck (FP) equation provide an appropriate framework for
the evolution of the heavy quark in the expanding
QGP heat bath. 

The system under study has two components. 
The equilibrium components, the  QGP, is 
assumed to be formed at a temperature $T_i$ 
at an initial time $\tau_i$ after the nuclear collisions. 
The QGP, due to its high internal pressure expands,
as a consequence it cools and reverts to hadronic 
phase at a temperature, $T_c$. The non-equilibrium
component, the heavy quarks produced due to the collision
of partons of the colliding nuclei has momentum distribution
determined by the perturbative QCD (pQCD), 
which  evolves due to their 
interaction with the expanding QGP background. 
The evolution of the heavy quark momentum distribution 
is governed by the FP equation.
The interaction of the heavy quarks with the QGP
is contained in the drag and diffusion coefficients.
The drag and diffusion coefficients are provided as inputs, which are,
in general, dependent on both temperature and momentum. 
The evolution of the temperature of the background 
QGP with time is governed by 
relativistic hydrodynamics. The solution of the FP equation
at the (phase) transition point for the charm and bottom quarks
gives the (quenched) momentum distribution of hadrons ($B$ and $D$ mesons) 
through fragmentation. The fragmentation of the initial momentum
distribution of the heavy quarks results in the unquenched 
momentum distribution of the $B$ and $D$ mesons.  The ratio
of the quenched to the unquenched $p_T$ distribution is the
nuclear suppression factor which is experimentally measured.
The quenching is due to the dragging of the heavy quark by QGP.
Hence the properties  of the QGP can be extracted
from the suppression factor. 
In contrast to earlier works where the momentum dependence of the drag 
coefficient were  ignored (or considered its value at low momentum)  
in the present work we emphasize on its momentum dependence.

The paper is organized as follows: In the next section we 
briefly describe the FP equation, in section III 
the drag coefficients for collisional and radiative 
processes have been discussed.  
In section IV the  initial conditions and the space
time evolution of the system have been described. Results
have been presented in section V and section VI has been dedicated   to
the summary and discussions.

\section{\bf The Fokker Planck Equation}
The Boltzmann transport equation describing non-equilibrium
statistical system reads:
\bea
\left[\frac{\partial}{\partial t}+\frac{\textbf{p}}{E}.\frac{\partial}{\partial \textbf{x}}
+\textbf{F}.\frac{\partial}{{\partial \textbf{p}}}
\right] f(\textbf{x},\textbf{p},t)=\left[\frac{\partial f}{\partial t}\right]_{collisions}
\eea
where \textbf{F} represents external forces acting on heavy quark, \textbf{p} and E denote 
three momentum and energy of the probe respectively.
In the absence of any external force in a uniform plasma and defining
\bea
f(\textbf{p},t)= \frac{1}{V} \int d^{3}\textbf{x} f(\textbf{x},\textbf{p},t)
\eea
we have for the normalized probability distribution in momentum space,
\be
\frac{\partial f(\textbf{p},t)}{\partial t}=\left[\frac{\partial f}{\partial t}\right]_{collisions}.
\ee
For the problem under consideration we need to evaluate the collision integral 
for a situation where one of the colliding partner is in thermal equilibrium.

To proceed in this direction, we define the collision rate of the heavy quark
with the thermal gluons as:
\be
w^{g}(\textbf{p},\textbf{k})= \gamma_{g}~\int \frac{d^{3}\textbf{q}}{(2\pi)^{3}}f'_{g}(\textbf{q})v_{\textbf{q,p}}\sigma^{g}_
{\textbf{p},\textbf{q}\ra \textbf{p}-\textbf{k},\textbf{q}+\textbf{k}}~~.
\label{rrate}
\ee
where $f_{g}$ is the thermal gluon distribution, $v_{\textbf{q},\textbf{p}}$ is the relative velocity 
between the two collision partners, $\gamma_{g}$ is the degeneracy of  gluon, and $\sigma^{g}$ is the 
cross section for  the heavy quark-gluon  elastic interaction. The collision
rate of the heavy quarks with the light quarks and anti-quarks can be evaluated in a way
similar to Eq.~\ref{rrate}.
In terms of the transition rates the collision integral of the Boltzmann transport equation can written as:
\be
\left[\frac{\partial f}{\partial t}\right]_{collisions}= \int d^{3}\textbf{k}[w(\textbf{p}+\textbf{k},\textbf{k})
f(\textbf{p}+\textbf{k})-w(\textbf{p},\textbf{k})f(\textbf{p})].
\ee
Using Landau approximation {\it i.e.} by expanding  
$w(\textbf{p}+\textbf{k},\textbf{k})$ in powers of  $\textbf{k}$
and keeping upto quadratic term,  
the Boltzmann transport equation can be written as
\be
\frac{\partial f}{\partial t}= \frac{\partial}{\partial p_{i}}\left[A_{i}(\textbf{p})f+\frac{\partial}{\partial p_{j}}[B_{ij}
(\textbf{p})f]\right]~~,
\label{landaukeq}
\ee
where the kernels are defined as
\be
A_{i}= \int d^{3}\textbf{k}w(\textbf{p},\textbf{k})k_{i}~~,
\label{eqdrag}
\ee
and
\be
B_{ij}= \frac{1}{2} \int d^{3}\textbf{k}w(\textbf{p},\textbf{k})k_{i}k_{j}.
\label{eqdiff}
\ee
Eq.~\ref{landaukeq} is a nonlinear integro-differential equation known as the Landau kinetic equation. 
For the problem under consideration one of the colliding partner is in equilibrium. In such situation
the distribution function which appears in $w$ can be replaced by thermal distribution. As a 
consequence Eq.\ref{landaukeq} becomes
a linear partial differential equation,
known as Fokker-Planck (FP) equation.  Assuming 
$A_{i}=p_i\gamma(p)$ and $B_{ij}=D(p)\delta_{ij}$, 
the FP equation can be written as:
\bea
\frac{\partial f}{\partial t}&=&C_{1}(p_{x},p_{y},t)\frac{\partial^{2}f}{\partial p_{x}^{2}}~+C_{2}(p_{x},p_{y},t)
\frac{\partial^{2}f}{\partial p_{y}^{2}}\nn\\&+&~C_{3}(p_{x},p_{y},t)\frac{\partial f}{\partial p_{x}}~+C_{4}(p_{x},p_{y},t)
\frac{\partial f}{\partial p_{y}}\nn\\&+&~C_{5}(p_{x},p_{y},t)f~+C_{6}(p_{x},p_{y},t).
\label{fpeqcartesian}~~.
\eea
where,
\bea
C_{1}&=& D\\C_{2}&=& D\\
C_{3}&=& \gamma ~p_{x}~+2~\frac{\partial D}{\partial p_{T}}~
\frac{p_{x}}{p_{T}}\\C_{4}&=& \gamma ~p_{y}~+2~\frac{\partial D}{\partial p_{T}}~\frac{p_{y}}{p_{T}}
\\C_{5}&=& 2~\gamma ~+\frac{\partial \gamma }{\partial p_{T}}~\frac{p_{x}^{2}}{p_{T}}~+
\frac{\partial \gamma }{\partial p_{T}}~\frac{p_{y}^{2}}{p_{T}}\\C_{6}&=& 0~~.
\eea
where the momentum, $\textbf{p}=(\textbf{p}_T,p_z)=(p_x,p_y,p_z)$. 
We numerically solve Eq.~\ref{fpeqcartesian} ~\cite{antia} with the boundary conditions:
$f(p_x,p_y,t)\ra 0$ for $p_x$,$p_y\ra \infty$ and  the initial (at time $t=\tau_i$)
momentum distribution of charm and bottom quarks are taken MNR code~\cite{MNR}.
It is evident from Eq.~\ref{fpeqcartesian} that with the momentum dependence 
transport
coefficients the FP equation becomes complicated.  It is possible to write 
down the solution of the FP equation in closed analytical form~\cite{rapp} in the special case of
momentum independent drag and diffusion coefficients. The computer code
used for the solution of Eq.~\ref{fpeqcartesian} has been used to reproduce 
the closed form analytical solution of Ref.~\cite{rapp}.

\section{Drag coefficient}
\subsection{Collisional process}
The drag coefficient, $\gamma_{coll}$ due to collisional process 
can be written as~\cite{svetitsky}:
\bea
\gamma_{coll}&=& \frac{1}{2E_{p}} \int \frac{d^{3}\textbf{q}}{(2\pi)^{3}2E_{q}}~\int \frac{d^{3}\textbf{q}'}{(2\pi)^{3}2E_{q'}}\nn\\
&\times&\int \frac{d^{3}\textbf{p}'}{(2\pi)^{3}2E_{p'}}~\frac{1}{\gamma_{Q}}\sum |M|^{2}
\nn\\&\times&(2\pi)^{4}\delta^{4}(p+q-p'-q')f'(\textbf{q})[1-\frac{\textbf{p}.\textbf{p}'}{p^{2}}]
\label{drag}
\eea
where $\textbf{p}'=\textbf{p}-\textbf{k}$ and $\textbf{q}'=\textbf{q}+\textbf{k}$.
The scattering matrix elements are given explicitly in Ref.~\cite{combridge}.
The integrations in Eq.~\ref{drag} has been performed using the 
standard techniques~\cite{svetitsky,dasetal}.  

\subsection{Radiative process}
The  drag coefficient due to the radiative process, $\gamma_{\mathrm rad}$ 
can be related to the energy loss as follows:
\be
-\left[\frac{dE}{dx}\right]_{rad}= \gamma_{rad}~p~~,
\ee
where $p$ is the momentum of the particle. 
We evaluate the radiative energy loss by using the 
following gluon spectrum~\cite{GB}:
\be
\frac{dn_{g}}{d\eta d^{2}k_{\bot}}= \frac{C_{A}\alpha_{s}}{\pi^{2}}~\frac{q_{\bot}^{2}}{k_{\bot}^{2}
[(\mathbf{q_{\bot}}-\mathbf{k_{\bot}})^{2}+m_{D}^{2}]}~~.
\label{gbmultiplicity}
\ee
where $k=(k_{0},k_{\bot},k_{3})$ is the four momentum,
$\eta=1/2\ln[(k_{0}+k_{3})/(k_{0}-k_{3})]$ is the rapidity of 
the emitted gluon and $q=(q_{0},q_{\bot},q_{3})$ is the four momentum
of the exchanged (propagator) gluon.  $C_{A}=3$ is
the Casimir invariant of the $SU(3)$ adjoint representation, $\alpha_{s}=g^{2}/4\pi$ is the strong coupling 
and $m_D$ is the Debye mass.  

The dead-cone effect is taken into account through the factor,
F~\cite{DK,rkellis}
\be
F= \frac{k_{\bot}^{2}}{k_{0}^{2}\theta_{0}^{2}+k_{\bot}^{2}}~~,
\ee
where $\theta_{0}=M/E$. 
The average energy loss per collision, $\epsilon$
is given by~\cite{changPL22,mustafathoma}
\bea
\epsilon&=& \langle n_{g}k_{0}\rangle= \int d\eta~d^{2}k_{\bot}~\frac{dn_{g}}{d\eta d^{2}k_{\bot}}\nn\\
&\times&k_{0}~\Theta (\tau_{sc} -\tau_{F})~\Theta (E-k_{\bot}\cosh\eta)F^2~~,
\label{avgenergy}
\eea
where the formation time of the emitted gluon~\cite{wang2}, 
$\tau_{F}=(C_{A}/2C_{2})~2\cosh\eta/k_{\bot}$, and $C_{A}/2C_{2}=N^{2}/(N^{2}-1)$
for quarks with $C_{2}=C_{F}=4/3$.  We perform the integrations in Eq.~\ref{avgenergy} 
and substitute the value of $\mathbf{q_{\bot}}$ by  its average value which is obtained as,
\be
\langle q^{2}_{\bot}\rangle= \frac{1}{\sigma}~\int^{(q^{max}_{\bot})^{2}}_{m^{2}_{D}} dq^{2}_{\bot}~\frac{d\sigma}
{dq^{2}_{\bot}}~q^{2}_{\bot}~~,
\ee
where,
\be
(q^{max}_{\bot})^{2}= \frac{s}{4}-\frac{M^{2}}{4}+\frac{M^{4}}{48pT}~\ln\left[\frac{M^{2}+6ET+6pT}{M^{2}+6ET-6pT}\right],
\ee
and $s\approx 6ET$, is the centre of mass energy squared for the scattering of a 
heavy quark with energy $E$ off the thermal partons at temperature $T$. 

The LPM effect is taken into account by introducing non-zero formation 
time  of the emitted gluon through the  first
theta function in Eq.~\ref{avgenergy}.
The scattering time scale, $\tau_{sc}$ can be estimated 
from the scattering rate ($\Lambda$) by using the relation,
$\tau_{sc}=\Lambda^{-1}$, where $\Lambda$ is given by:
\bea
\Lambda&=& \frac{1}{2E_{p}} \int \frac{d^{3}\textbf{q}}{(2\pi)^{3}2E_{q}}~\int \frac{d^{3}\textbf{q}'}{(2\pi)^{3}2E_{q'}}\nn\\
&\times&\int \frac{d^{3}\textbf{p}'}{(2\pi)^{3}2E_{p'}}~\frac{1}{\gamma_{Q}}\sum |M|^{2}\nn\\&\times&(2\pi)^{4}\delta^{4}
(p+q-p'-q')f(q')
\label{rate}
\eea
Note that deletion of $k_i$ from the expression for $A_i$ (Eq.~\ref{eqdrag})
gives the scattering rate, $\Lambda$.
Knowing $\epsilon$ and $\Lambda$
the energy loss of heavy quark can be expressed as:
\be
-\left[\frac{dE}{dx}\right]_{rad}= \Lambda \epsilon~~.
\label{elosshq}
\ee
The drag due to radiative loss can now be estimated using Eqs.~\ref{avgenergy} and \ref{elosshq}.
The effective drag due to collisional and radiative processes is obtained as:
$\gamma= \gamma_{coll}+\gamma_{rad}$. 
It should be mentioned that
the radiative process is not fully independent of the collisional one.
Therefore, the collisional and radiative transport coefficient may 
not be  added to obtain the effective one.
In absence of any rigorous method, however, we add them up.  Since 
the radiative loss is larger than the collisional one, therefore,
this  approximation may not be treated as unreasonable.

A comment on the effects of Bose enhancement (BE) of the 
gluons and Pauli blocking (PB) of the  
quarks \& antiquarks in the final channels 
on the drag is in order here.
We have found that the average change due 
to BE and PB effects is about $\sim 6\%$  for $p<5$ GeV,
at higher $p$ the change is negligible for the temperature domain
under consideration.  The effects of BE and PB  is inconsequential on the
nuclear suppression factor evaluated later.
The diffusion coefficient has been obtained  from the drag
coefficients by using the Einstein's relation, $D=\gamma M T$,
where $M$ is the mass of the heavy quark. It is important
to mention at this point that the Einstein's relation is 
valid in the non-relativistic domain, use of this relation
leads to underestimation of the value of $D$
in the relativistic realm ~\cite{moore}. 
In our calculation the temperature dependence of 
the $\alpha_s$~\cite{kaczmarek} has been taken into account.


\section{\bf Initial conditions}
In order to solve Eq.~\ref{fpeqcartesian} we supply the initial distribution functions, $f_{in}(p_{T},t)$ for charm and bottom
quarks from the well known MNR code~\cite{MNR}. The initial temperature,$T_{i}$ and the initial thermalization time,
$\tau_{i}$ for the background QGP expected to be formed at RHIC and LHC can be constrained  to the total
multiplicity as follows:
\be
T^{3}_{i}\tau_{i}\approx \frac{2\pi^{4}}{45\zeta(3)}~\frac{1}{4a_{eff}}~\frac{1}{\pi R^{2}_{A}}~\frac{dN}{dy}~~,
\ee
where $R_{A}$ is the radius of the system, $\zeta(3)$ is the Riemann zeta function, $a_{eff}=\pi^{2}g_{eff}/90$,
and $g_{eff}(=2\times8+7\times2\times2\times3\times N_{F}/8)$ is the degeneracy of quarks and gluons in QGP and $N_{F}$
is the number of flavours. The value of the transition temperature, $T_{c}$ has been taken to be 175 MeV. We have used the 
boost invariant model of relativistic hydrodynamics proposed by 
Bjorken~\cite{jdbjorken} for the
space time evolution of the expanding QGP back ground.
The geometry of the collision and the space time evolution has been described
in detail earlier, therefore, we avoid
further discussions here and refer to our previous work~\cite{dasetal} for
details.  
The value of $T_i$ and $\tau_i$ for the QGP
fireball are taken as $T_i=300$ MeV
and $\tau_i=0.5$ fm/c. The corresponding quantities for LHC are 
$T_i=550$ MeV and $\tau_i=0.1$ fm/c. The pressure ($P$)-energy 
density($\epsilon$) relation for the QGP has been taken as $P=\epsilon/3$.

\begin{figure}[h]
\begin{center}
\includegraphics[scale=0.43]{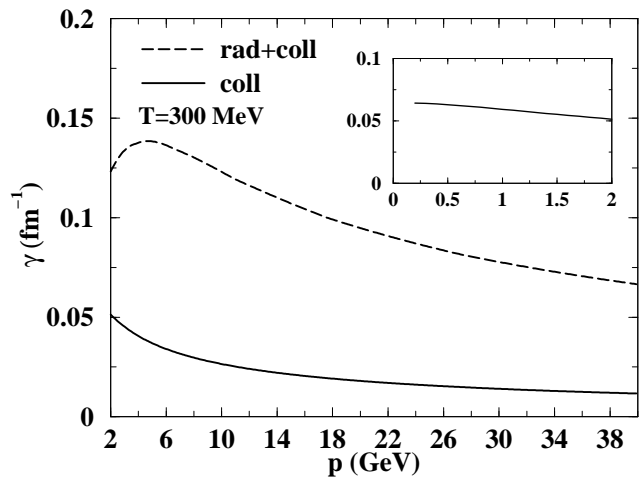}
\caption{Drag coefficients of charm assuming running 
strong coupling,  $\alpha_{s}(T)$ and temperature dependent
Debye screening mass, $m_{D}(T)$ due to its interaction
with thermal gluons, quarks, and antiquarks.
Inset: The variation of drag with $p$ in the  low domain.}
\label{fig1}
\end{center}
\end{figure}
\begin{figure}[h]
\begin{center}
\includegraphics[scale=0.43]{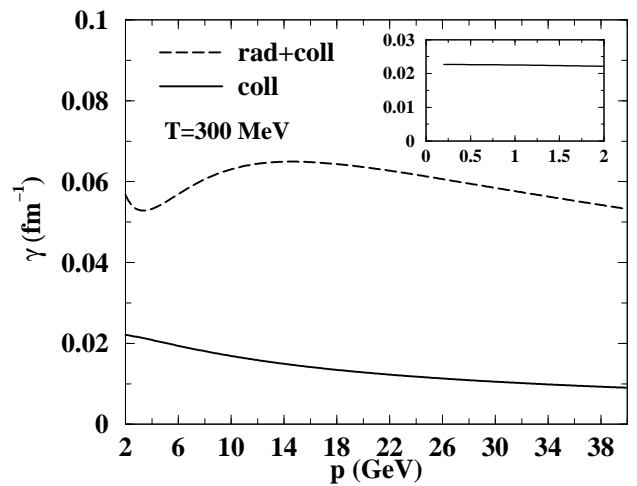}
\caption{Drag coefficients of bottom assuming running 
strong coupling, $\alpha_{s}(T)$ and temperature dependent Debye
screening mass, $m_{D}(T)$ due to its interaction with
thermal  gluons, quarks, and antiquarks.
Inset: The variation of drag with $p$ in the  low domain.}
\label{fig2}
\end{center}
\end{figure}

\section{Results}
With the formalism described  and the inputs mentioned
above, we present the results now. 
The momentum dependence of the drag coefficient ($\gamma$)  of charm quark
propagating through QGP has been displayed in Fig.~\ref{fig1} 
for $T=300$ MeV.
The variation of $\gamma$ with $p$ is non-negligible both 
for radiative as well as collisional processes.
The drag coefficients, $\gamma$ is related to the 
energy loss as: $\gamma\sim (1/p)\Delta E/\Delta x$ 
which in the high energy limit ($E\sim p$) varies very
slowly with $E$ (Fig.~\ref{fig1}), a behaviour very similar to the
one observed in the fractional energy loss as a function 
of the initial energy  
of light quark jets~\cite{gyqin}. 
The value of $\gamma$ due to collisional 
processes at $p=5$ GeV is about 0.036 fm$^{-1}$ which reduces to a value of
0.018 fm$^{-1}$ at $p=10$ GeV. Such a variation,
which was neglected earlier, will have crucial consequences
on the nuclear suppression factor,  $R_{\mathrm AA}$ for the 
charm and bottom quarks.  
In the inset of Fig.~\ref{fig1} 
we display the drag coefficient for collisional
process for the low momentum domain. In the limited momentum range 
the drag remains almost independent of momenta. This constant values of 
drag has been used earlier in the FP equation and subsequently 
the solution  was used  in estimating the nuclear suppression
factor.  Since the value of $\gamma$ reduces with $p$ 
one will overestimate the suppression by taking its value at low $p$.
For the low momentum region we use $\gamma=\gamma_{coll}$. 
In Fig.~\ref{fig2} we depict the drag
coefficients of a bottom quark  which shows slower 
(compared to charm) variation with momentum.
The drag coefficient for collisional loss has been shown 
in the inset  of Fig.~\ref{fig2} for low momentum domain. 

\begin{figure}[h]
\begin{center}
\includegraphics[scale=0.43]{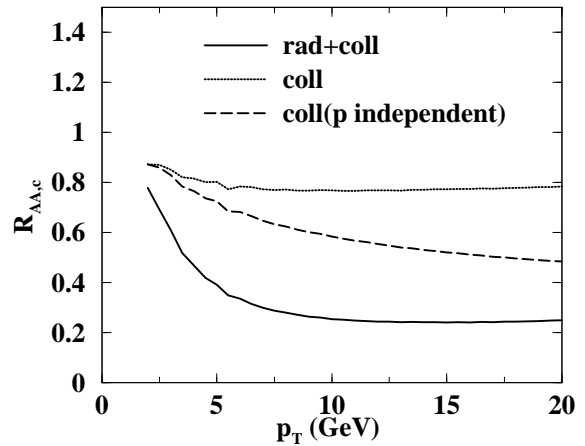}
\caption{Suppression of momentum of charm quarks in QGP as a function $p_{T}$ 
}
\label{fig3}
\end{center}
\end{figure}

\begin{figure}[h]
\begin{center}
\includegraphics[scale=0.43]{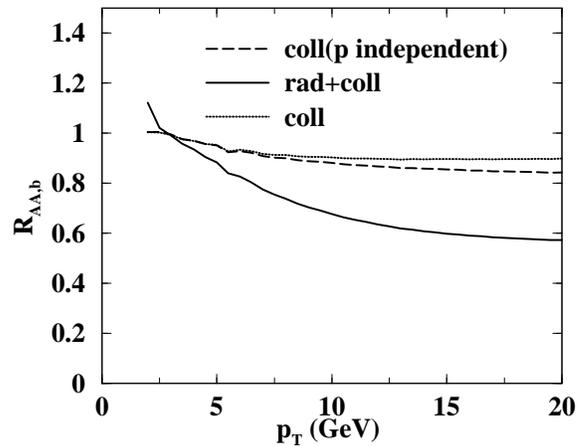}
\caption{Suppression of momentum of bottom 
quarks in QGP as a function $p_{T}$ 
}
\label{fig4}
\end{center}
\end{figure}

The  suppression of both charm and bottom quarks~\cite{dasetal}
(before fragmentation to 
hadrons) are plotted against $p_T$ in Figs.~\ref{fig3} and ~\ref{fig4}
respectively.  We note that if one takes the drag to be 
momentum independent (or more precisely takes the
value of $\gamma$ at low $p$ and extends it upto very 
high $p$) then the drag due to collisional process 
causes about $50\%$ suppression (dashed line).   
However, if we take into account 
the variation of $\gamma$ with $p$ obtained from 
pQCD calculation  then about  $20\%$ of suppression
can be achieved, {\it i.e.}  the contribution from the
collisional loss becomes smaller with the momentum dependent drag.
Therefore, the observed large suppression of the 
heavy quarks at RHIC is predominantly due to radiative loss.
In fact, the inclusion of the radiative processes increases the
suppression to about $75\%$. This can be understood from the
fact that the drag due to the radiative loss
is large. The suppression of the bottom quark is much less
because of the smaller values of drag and 
initial harder momentum distribution (Fig.~\ref{fig4}).

\begin{figure}[h]
\begin{center}
\includegraphics[scale=0.43]{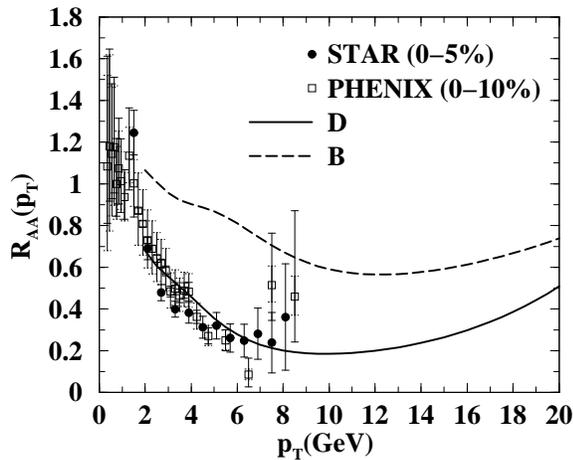}
\caption{$R_{AA}$ as a function of  $p_T$ for $D$ and $B$ mesons  at RHIC.
Experimental data taken from ~\cite{stare} and ~\cite{phenixe}.}
\label{fig5}
\end{center}
\end{figure}

The hadronization of charm and bottom quarks
to $D$ and $B$ mesons respectively are done by using
Peterson fragmentation function~\cite{peterson}.
The variation of the nuclear suppression factor, $R_{AA}$
~\cite{dasetal} with $p_T$ of the
$D$ and $B$ mesons has been displayed in Fig.~\ref{fig5}
for RHIC initial condition ($T_i=300$ MeV). 
The suppression for bottom is much less for the 
reasons mentioned earlier.
The theoretical results show a slight upward 
trend for $p_T$ above 10 GeV both for mesons containing 
charm and bottom quarks.  Similar trend has recently been experimentally
observed for light mesons at LHC energy~\cite{CMS}.
This may originate from the fact that 
the drag (and hence the quenching) for charm and 
bottom quarks are less at higher momentum. 

\begin{figure}[h]
\begin{center}
\includegraphics[scale=0.43]{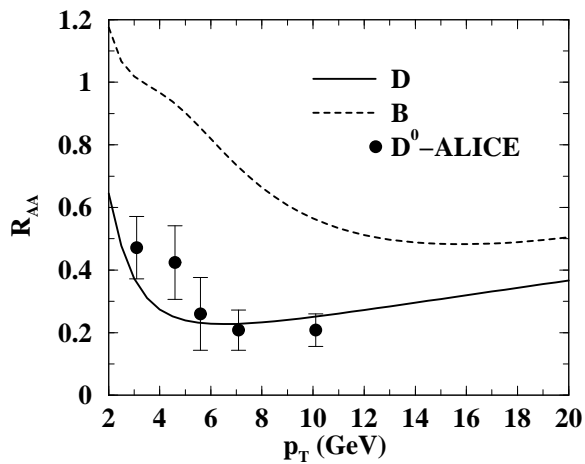}
\caption{$R_{AA}$ as a function of  $p_T$ for $D$ and $B$ mesons  at LHC.
Experimental data taken from ~\cite{ALICE}.}
\label{fig6}
\end{center}
\end{figure}

The same formalism is extended to evaluate the nuclear suppression 
factor, $R_{AA}$ both for charm and bottom at LHC energy. Result has been 
compared with the recent ALICE data(Ref.~\cite{ALICE}) in Fig.~\ref{fig6}.
The data is reproduced well by assuming formation of QGP 
at an initial temperature $\sim 550$ MeV after Pb+Pb collisions at 
$\sqrt{s_{\mathrm NN}}=2.76$ TeV.  

\section{\bf Summary and discussions}
The temperature and momentum dependence of 
drag and diffusion coefficients of heavy quarks interacting with
the thermal partonic medium 
have been evaluated by using the elastic and inelastic interactions.
We have employed both the dead-cone and the LPM effects
in the calculation for the inelastic processes.
The initial $p_T$ distributions for charm and bottom are taken from 
MNR code~\cite{MNR}. 
The FP equation has been solved with the momentum dependent 
transport coefficients and subsequently the  
nuclear suppression factors, $R_{AA}$ have been calculated for 
$D$ and $B$ mesons for RHIC and LHC conditions.  The momentum dependence of the
drag coefficient is found to be crucial in reproducing the trend in the $p_T$
dependence of the experimental data. It has been seen that the 
radiative loss plays more dominant role than the
collisional  process. 

\section*{\bf ACKNOWLEDGMENT}
We are grateful to Matteo Cacciari for providing us the heavy
quarks transverse momentum distribution and also for useful discussions.
JA and SKD are partially supported by DAE-BRNS project
Sanction No.  2005/21/5-BRNS/245.

\end{document}